\documentclass{PoS}
\usepackage{cancel}
\usepackage{amsmath}
\title{An overview of light-front holography}

\ShortTitle{An overview of LFH}

\author{\speaker{Ruben Sandapen}\thanks{I thank the organizers of Light Cone 2019 for their invitation and a very successful conference. I also thank the Natural Sciences and Engineering Research Council (NSERC) of Canada for funding (Discovery Grant No. SAPIN-2017-00031)}\\
Acadia University\\ 15 University Avenue \\Wolfville NS Canada \\B4P 2R6\\
        E-mail: ruben.sandapen@acadiau.ca}
\abstract{Light-front holography offers a successful first semiclassical approximation to hadronic spectroscopy and dynamics. We review its underlying assumptions, its remarkable predictions as well as attempts to go beyond the semiclassical approximation in order to the describe a wide range of data with a universal AdS/QCD mass scale.}

\FullConference{Light Cone 2019 - QCD on the light cone: from hadrons to heavy ions - LC2019\\
		16-20 September 2019\\
		Ecole Polytechnique, Palaiseau, France}

\begin{document}

\section{Conformal symmetry in QCD}
In the chiral limit and at tree-level, the QCD Lagrangian is invariant under conformal transformations. This underlying conformal symmetry of QCD motivates the search for its gravity dual according to Maldacena's equation (AdS=CFT) which refers to an equivalence 
between weakly-coupled gravitational theories in higher dimensional anti-de Sitter (AdS) space and strongly-coupled conformal field theories (CFT) in physical spacetime \cite{Maldacena:1997re}.  In QCD, conformal invariance is explicitly broken by non-zero quark masses ($m_f$) and also  beyond tree-level where renormalization generates a scheme-dependent mass scale: $\Lambda_{\mathrm{QCD}}$.  Since the Higgs-generated quark masses are much smaller than hadron masses, it is usually stated that $\Lambda_{\mathrm{QCD}}$  sets the scale for hadron masses. 

Yet, there is another way to generate a scale for the hadron masses even within tree-level, massless QCD. A mass scale can appear in the Hamiltonian and equations of motion while the underlying action remains conformally invariant as was first shown by de Alfaro, Fubini and Furlan (dAFF) in conformal QM \cite{deAlfaro:1976vlx}. To apply the dAFF procedure in QCD, we need to reduce the strongly-coupled multi-parton bound state problem in QCD to a two-parton QM-like bound state problem. This is possible in light-front QCD where \cite{Brodsky:1997de}
\begin{equation}
H_{\mathrm{LF}}^{\mathrm{QCD}} |\Psi(P) \rangle=M^2 |\Psi(P) \rangle
\label{Heisenberg}	
\end{equation}
can be reduced to a Schr\"odinger-like equation for the lowest quark-antiquark Fock state:
\begin{equation}
	\left(\frac{\mathbf{k}_{\perp}^2 + m_f^2}{x(1-x)} + U_{\mathrm{eff}}(x,\mathbf{k}_{\perp}) \right) \Psi(x,\mathbf{k_{\perp}},h,\bar{h})= M^2 \Psi(x,\mathbf{k_{\perp}},h,\bar{h})  
	\label{LFSE}
\end{equation}	
where $\Psi(x,\mathbf{k_{\perp}},h, \bar{h})=\langle q \bar{q}:x,\mathbf{k_{\perp}},h,\bar{h}|\Psi\rangle$ is the valence wavefunction, with $x$ being the light-front momentum fraction carried by the quark, $\mathbf{k_{\perp}}$ is its transverse momentum and $h$ its helicity ($\bar{h}$ is the helicity of the antiquark).  At this point, there are no approximations and the effective potential $U_{\mathrm{eff}}$, which describes the quark-antiquark interaction as well as the effects of higher Fock states on the lowest Fock state, is essentially unknown. To make further progress, two assumptions are necessary.  First that the helicity dependence of the wavefunction is non-dynamical, i.e. 

\begin{equation}
	\Psi(x,\mathbf{k_{\perp}}, h, \bar{h}) \to  \Psi(x,\mathbf{k_{\perp}}) S_{h \bar{h}}
	\end{equation}
	where $S_{h \bar{h}}$ is momentum-independent. For example, $S_{h \bar{h}} \propto h \delta_{h,-\bar{h}}$ for the pion. Second that, in the chiral limit, the confinement potential is a function of the invariant quark-antiquark mass squared, i.e.
\begin{equation}
	\mathcal{M}^2|_{m_f \to 0} = \frac{\mathbf{k}_{\perp}^2 + \cancelto{0}{m_f}^2}{x(1-x)} \;.
\label{Invariant-mass}
\end{equation}
Then, the wavefunction can be factorized as
	\begin{equation}
	\Psi(x,\zeta,\varphi)=\frac{\phi(\zeta)}{\sqrt{2\pi \zeta}} X(x) e^{iL\varphi} 
	\label{hLFWF}
	\end{equation}
where $\mathbf{\zeta}^2 = x(1-x) \mathbf{b}^2_{\perp}$ is the 2-dimensional Fourier transform of  $\mathcal{M}^2|_{m_f \to 0}$. Eq. \eqref{LFSE} then becomes
	\begin{equation}
			\left(-\frac{\mathrm{d}^2}{\mathrm{d}\zeta^2}-\frac{1-4L^2}{4\zeta^2} + U_{\mathrm{eff}}(\zeta) \right) \phi(\zeta)=M^2 \phi(\zeta) \;.
	\label{hSE}
	\end{equation}
If $U_{\mathrm{eff}}=0$, the Hamiltonian in Eq. \eqref{hSE} maps onto the conformal QM Hamiltonian considered by dAFF \cite{deAlfaro:1976vlx}. Then, following dAFF, taking
\begin{equation}
U_{\mathrm{eff}}(\zeta)=\kappa^4 \zeta^2 
\label{quadratic}
\end{equation}
introduces a mass scale $\kappa$ in the Hamiltonian without spoiling the conformal invariance of the underlying action.  It is remarkable that the quadratic form of the confinement potential is uniquely fixed by the dAFF mechanism of conformal symmetry breaking (at this point, there is no reference to AdS space). For this reason, it is important to recap the underlying assumptions leading to Eq. \eqref{quadratic}: no quantum loops, zero quark masses and the identification of $\zeta$ as the variable on which the confinement potential depends. These three assumptions are encapsulated in the so-called semiclassical approximation \cite{Brodsky:2014yha}. Here, we consider the latter to also include the assumption of a non-dynamical spin wavefunction.

\section{Light-front holography}
In the semiclassical approximation, light-front QCD possesses a gravity dual. With $\zeta \leftrightarrow z$, where $z$ is the $5^{\mathrm{th}}$ dimension of $\mathrm{AdS}_5$, and $L^2 \leftrightarrow (\mu R)^2 + (2-J)^2$ where $\mu$ is a 5-dimensional mass parameter and $R$ is the radius of curvature of $\mathrm{AdS}_5$, Eq. \eqref{hSE} maps onto the classical wave equation for freely propagating spin-$J$ string modes in AdS  \cite{deTeramond:2008ht}. The confining QCD potential is then determined by the dilaton field which breaks conformal invariance in AdS: \cite{Brodsky:2014yha}
	\begin{equation}
	U_{\mathrm{eff}}(\zeta)= \frac{1}{2} \varphi^{\prime \prime}(z) + \frac{1}{4} \varphi^{\prime}(z)^2 + \frac{2J-3}{2 z} \varphi^{\prime}(z) \;.
	\label{dilaton-potential}
	\end{equation}	
To recover the quadratic effective potential dictated by the dAFF mechanism, the dilation field must be chosen to be quadratic, i.e. $\varphi(z)=\kappa^2 z^2$. Eq. \eqref{dilaton-potential} then yields 
\begin{equation}
		U_{\mathrm{eff}}(\zeta)=\kappa^4 \zeta^2 + 2 \kappa^2 (J-1)	\;.
	\label{hUeff}	
\end{equation}
As can be seen, while the quadratic form of the potential is fixed by the dAFF mechanism, the constant spin term emerges from the mapping to AdS space \cite{Brodsky:2013ar}. The mass scale $\kappa$ which appears in both the dilaton field and the confinement potential is called the AdS/QCD mass scale. With its potential given by Eq. \eqref{hUeff}, Eq. \eqref{hSE} is referred to as the holographic Schr\"odinger Equation. Its eigenvalues are:
\begin{equation}
 	M_{n,L,S}^2= 4\kappa^2 \left(n+L +\frac{S}{2}\right)\;
 	\label{mass-Regge}
 \end{equation}
where $J=L+S$ while its eigenfunctions,
 \begin{equation}
 	\phi_{nL}(\zeta)= \kappa^{1+L} \sqrt{\frac{2 n !}{(n+L)!}} \zeta^{1/2+L} \exp{\left(\frac{-\kappa^2 \zeta^2}{2}\right)}  ~ L_n^L(\kappa^2 \zeta^2)\;,
 \label{phi-zeta}
 \end{equation}
are normalized so that
\begin{equation}
	\sum_{h,\bar{h}} \int \mathrm{d}^2 \mathbf{b} \mathrm{d} x |\Psi_{h,\bar{h}}(x, \mathbf{\zeta})|^2 =1 \;.
	\label{norm}
\end{equation}
Eq. \eqref{norm} embodies the assumption that the meson consists only of the leading quark-antiquark Fock state.  

The first striking prediction is that the lowest lying bound state, with quantum numbers $n=L=S=0$, is massless: $M^2=0$.  This state is naturally identified with the pion since the  pion mass is expected to vanish in chiral limit. To completely specify the light-front wavefunction, we need to fix $X(x)$. This is done by matching the electromagnetic or gravitational form factor of composite states in physical spacetime and in $\mathrm{AdS}_5$, resulting in $X(x)=\sqrt{x(1-x)}$ \cite{Brodsky:2014yha}. 

Up to now, we have considered only quark-antiquark states, i.e. mesons. To make this explicit, let us write $U_{\mathrm{eff}} (\zeta) = 	U_{M}(\zeta,\kappa^2)$. The supersymmetrization of the holographic Schr\"odinger Equation unifies mesons and baryons (considered as diquark-quark systems) as superpartners \cite{Dosch:2015nwa}.  Just like in the supersymmetric formulation of ordinary QM \cite{Witten:1981nf}, the partner potential to a harmonic oscillator potential is a shifted harmonic oscillator potential:
\begin{equation}
	U_B (\zeta, \kappa^2)= U_M(\zeta, \kappa^2) + 2 \kappa^2  
\label{UB-UM}
\end{equation}
where our notation has anticipated the fact that Eq. \eqref{UB-UM} is the potential for baryons. Explicitly
\begin{equation}
	U_B(\zeta, \kappa^2)=\kappa^4 \zeta^2 + 2 \kappa^2 ((L_B+1) + S_D -1)
\label{UB}
\end{equation}
where $S_D$ is the lowest possible spin of the diquark in the baryon. Eq. \eqref{mass-Regge} then generalizes to
\begin{equation}
	M_{n,f}^2=4\kappa^2 \left(n+f+\frac{1}{2}\right)
	\label{mesons-baryons}
\end{equation}
where $f=L_B +1/2(1 + S_D)=L_M-1/2(1-S_M)$. This implies that mesons and baryons differing by only one unit of orbital angular momentum (i.e. with $L_M=L_B+1$ and $S_M=S_D$) lie on identical Regge trajectories. This degeneracy, although not exact, is indeed seen in spectroscopic data. Note that once Eq. \eqref{hUeff} is fixed by light-front holography, Eq. \eqref{UB} is completely fixed by supersymmetry without further reference to $\mathrm{AdS}_5$. On the other hand, light-front holography tells us that a baryon has equal probability of being in the positive and negative chirality states, with orbital angular momentum $L_B$ and $L_B+1$ respectively \cite{Brodsky:2014yha}. Therefore, in addition to mesons, there are second bosonic superpartners to baryons with quantum numbers $(L_B,S_D)$. These are interpreted as tetraquarks (considered as diquark-antidiquark systems) with quantum numbers $(L_T=L_B, S_T=S_D)$ \cite{Nielsen:2018uyn}. Note that the massless pion does not have a baryon superpartner.

The holographic wavefunction, Eq. \eqref{hLFWF}, is a first approximation to the meson light-front wavefunction which is a crucial input to predict various observables like charge radii, decay constants, diffractive cross-sections as well as non-perturbative quantities like Form Factors (FF), Parton Distribution Functions (PDFs), Distribution Amplitudes (DAs), Generalized Parton Distributions (GPDs) and Transverse Momentum Distributions (TMDs). In what follows, we shall focus on predictions using the holographic meson wavefunction. For other applications of light-front holography, see \cite{Brodsky:2019npe} and references therein. 

\section{Beyond the semiclassical approximation}
The superconformal symmetry leading to Eq. \eqref{mesons-baryons} is broken by non-zero quark masses. Light quark masses can be treated as a small perturbation shifting the hadronic masses by \cite{Brodsky:2014yha}
\begin{equation}
	\Delta M^2 [m_1,...m_n]=\kappa^4 \frac{\partial \ln F}{\partial \kappa^2}
\label{quark-masses}
\end{equation}
where
\begin{equation}
	F[\kappa^2]=\int_0^1 \mathrm{d} x_1 ... \mathrm{d} x_n \delta \left(\sum_i^n x_i-1\right)\exp{\left(-\frac{1}{\kappa^2} \sum_i^n \frac{m_i^2}{x_i} \right)}
\end{equation}
where $n=2$ for mesons and $n=3$ for baryons. This allows us to fix the $u/d$ quark masses using the fact that the physical pion mass $M_\pi=\Delta M (m_{u/d})=140$ MeV. This gives $m_{u/d}=46$ MeV. The strange quark mass can then be fixed using $M_K=\Delta M (m_{u/d}, m_s)=494$ MeV, yielding $m_s=357$ MeV. A global fit to all light meson and baryon spectroscopic data yields $\kappa=523 \pm 24$ MeV \cite{Brodsky:2016rvj} which we refer to as the universal AdS/QCD scale. Being scheme-independent, $\kappa$ can be viewed as being more fundamental than $\Lambda_{\mathrm{QCD}}$ and indeed the latter can be predicted from the former \cite{Deur:2016opc}. While $\kappa$ is universal for light hadrons, it depends on the hadron mass (as expected from Heavy Quark Effective Theory) for  hadrons containing at least one heavy quark \cite{Dosch:2016zdv}. The latter observation was already noted in earlier research \cite{Gutsche:2014oua,Branz:2010ub} even though the more recent work \cite{Nielsen:2018ytt,Nielsen:2018uyn} seem to provide a more economical way (i.e. with less free parameters) to account for heavy quarks. In particular, the only correction to the heavy hadron mass spectrum is that given by Eq. \eqref{quark-masses}, with no short-distance corrections as in \cite{Gutsche:2014oua,Branz:2010ub}, despite the fact that conformal symmetry is strongly broken by heavy quarks.

Eq. \eqref{phi-zeta} tells us that the pion and the $\rho$ meson have the same holographic wavefunction since for both mesons, $n=1$ and $L=0$. With a universal $\kappa$, this leads to degenerate decay constants for the two mesons, a prediction that is in contradiction with experiment.  It has been argued that the pion is a special case since it does not fall on a Regge trajectory \cite{Vega:2009zb} and, indeed in previous work \cite{Vega:2009zb,Branz:2010ub,Swarnkar:2015osa}, a different $\kappa$ and/or normalization condition (instead of Eq. \eqref{norm}) were used only for the pion. Nevertheless, these approaches cannot accommodate a non-zero holographic pion Boer-Mulders TMD which results from the overlap between the $L=0$ and $L=1$ components of the pion wavefunction \cite{Ahmady:2019yvo}. These shortcomings can be overcome by taking into account dynamical spin effects.\footnote{Only the unpolarized pion TMD can be predicted with the holographic wavefunction with no dynamical spin \cite{Bacchetta:2017vzh}.}

The assumption is that bound state effects are fully captured by the holographic wavefunction so that the spin wavefunction is that of a point-like meson coupling to a quark-antiquark pair. Hence, 
\begin{equation}
	\Psi(x, \mathbf{k}, h, \bar{h})\to \Psi(x, \mathbf{k}) S_{h, \bar{h}}(x, \mathbf{k})
\end{equation}
where $\Psi(x, \mathbf{k})$ is the holographic wavefunction in momentum space and 
 \begin{equation}
	S_{h, \bar{h}}(x, \mathbf{k})= \frac{\bar{v}_{\bar{h}}((1-x)P^+,-\mathbf{k})}{\sqrt{1-x}} \Gamma \frac{u_{h}(xP^+,\mathbf{k})}{\sqrt{x}} 
\label{vector}
\end{equation}
is the (point-like) meson-quark-antiquark vertex in light-front perturbation theory. In general, the Dirac structure between the light-front spinors is given $\Gamma=\epsilon \cdot \gamma$ for vector mesons \cite{Forshaw:2012im,Ahmady:2016ujw} and $\Gamma = (P \cdot \gamma + B M_P) \gamma^5$ for pseudoscalar mesons \cite{Ahmady:2018muv,Ahmady:2016ufq}. Here $\epsilon^\mu$ is the polarization $4$-vector of the vector meson, $P^\mu (M_P)$ is the $4$-momentum (mass) of the pseudoscalar meson and $B$ is a dimensionless free parameter.  It was shown in \cite{Forshaw:2012im} that, with a spin structure given by Eq. \eqref{vector}, the HERA data on diffractive $\rho$ meson electroproduction prefer the holographic wavefunction with $\kappa = 540$ MeV (consistent with the universal value). Diffractive $\phi$ electroproduction can also be successfully predicted \cite{Ahmady:2016ujw} with the same value of $\kappa$ and the best simultaneous description of diffractive $\rho$ and $\phi$ electroproduction data is achieved with $[m_{u/d},m_s]=[46,140]$ MeV. The holographic DAs for $\rho$, $\phi$ and $K^*$ have also been predicted and used to compute various observables in B decays: see \cite{Ahmady:2020xjx} and the references therein. For pseudoscalar mesons, Ref. \cite{Ahmady:2016ufq} reports an excellent description of the pion data with the universal $\kappa$ and $[m_{u/d},m_s]=[330,500]$ MeV. Ref. \cite{Ahmady:2018muv} cautions against using a universal $B$b for the pseudoscalar meson nonet: while the pion data prefer  $B \ge 1$, the available kaon data set prefers $B=0$ while the situation is less clear for the $\eta/\eta^\prime$ system (even though the meson-to-photon transition FF data prefer $B \ge 1$). Further support for the spin-improved wavefunction  comes from the configurational entropy analysis of \cite{Karapetyan:2018yhm} and lattice data \cite{Ahmady:2020ynt}. Alternative ansatzes for the spin structure of mesons have been proposed in \cite{Chang:2018aut}.

\section{Summary}
Superconformal light-front holography successfully predicts the main features of hadronic 
spectroscopy especially the near mass degeneracy of mesons and baryons differing by one unit of orbital angular momentum. Non-zero quark masses and dynamical spin effects can be accounted for \textit{\`a posteriori} leading to a successful description of a wide set of light meson data with a universal AdS/QCD scale which governs both confinement and spectroscopy. 

\bibliographystyle{JHEP}
\bibliography{RSandapenLC2019.bib}

\end{document}